\documentstyle[titlepage,fleqn,12pt]{article}
\begin{document}
\begin{titlepage}
\title{Growth model of Au films on Ru(001)}
\author{E. Canessa and A. Calmetta \\
\\
ICTP-International Centre for Theoretical Physics\\
Condensed Matter Group\\
P.O. Box 586, 34100 Trieste\\
ITALY}
\date{}

{\baselineskip=17pt

{}.
\vspace{2cm}

\begin{center}
{\bf Abstract}
\end{center}

In an attempt to find generic features on the fractal growth of Au
films deposited on Ru(001), a simple simulation model based on irreversible
diffusion-limited aggregation (DLA) is discussed.  Highly irregular
two-dimensional dentritic islands of Au particles that gradually grow on a
larger host lattice of Ru particles and have fractal dimension
$d_{f}\approx 1.70$ each, are generated via a multiple {\em had-hoc}
version of the DLA algorithm for single aggregates.
Annealing effects on the islands morphology
are reproduced assuming different sticking probabilities at
nearest-neighbour lattice sites of Au films on Ru(001).  Using simulation
data, islands growth are described in analogy to diffusion-limited,
precipitate growth with soft impingement of precipities.  This leads to
analyse thin film island growth kinetics in such fractal systems and to
predict a main peak in scattering intensity patterns due to interisland
interference.

}

\maketitle
\end{titlepage}
\baselineskip=22pt
\parskip=0pt

The two-dimensional (2D) thin film growth on single-crystal surfaces
is being a subject of considerable interest \cite{Lei92,Hw91}.
Particularly so are the recent experimental investigations on
Au thin-films deposited on Ru(001) which have revealed novel
and intriguing phenomena \cite{Hw91}.  Upon deposition at room
temperature, it has been found that highly dendritic Au islands
(one layer thickness and width of approximately $100\; \AA$)
nucleate on several larger Ru terraces to form
fractal aggregates with an average fractal dimension
$d_{f}\approx 1.72 \pm 0.07$ each.  Hence, such denditric island shapes
are believed to be due to diffusion-limited
aggregation (DLA) mechanism \cite{Wi81}.  Upon annealing to $650\; K$,
the islands are seen to collapse into compact forms.

Despite theoretical efforts over the last decade in the area
of single fractal structures (see {\em e.g.} \cite{Ma82,Vi89,Pi89}),
the natural extension of phenomena to the `simultaneous' growth
of 2D dentritic islands diffusing on multiple terraces
is hampered by the complexity of this problem.  The presence of
several moving interfaces, whose local velocities may be in turn
determined by the normal gradient of the Laplace field on each
island \cite{Co91}, is not readily amenable to modern
theories \cite{Pi89,Ni84} especially so because of the complex
boundary conditions.
An alternative to understand these aggregation phenomena
consists of applying the DLA simulation approach of Witten and Sander
\cite{Wi81} to model the gradual grow of aggregates by the succesive addition
of random walkers to the perimeter sites of (many) forming islands.
This is the aim of this letter.

In an attempt to find generic features on the fractal growth of Au
films deposited on Ru(001), we shall report the results of a simple
simulation model based on irreversible DLA.
Via a multiple {\em had-hoc} version of the DLA algorithm, we generate
(13) highly irregular two-dimensional
dentritic islands of (Au) particles that gradually grow on a larger host
lattice of (Ru) particles and, even more important, present fractal dimension
$d_{f}\approx 1.70$ each.  We mimic
the annealing effects on the (several) clusters morphology
by assuming different sticking probabilities at
nearest-neighbour lattice sites of Au films on Ru(001) \cite{Me83}.
{}From the results of simulations, which are shown to display relevant
experimental features, we predict a main peak in scattering intensity
patterns due to the fact that the average edge-to-edge island spacings
are less than or equal the average island diameters.

The multiple DLA simulation model we use starts by locating
13 seed particles at the corners, edges and (sheared) centers of
two concentric squares which in turn are contained on a third, larger 2D
square lattice.  The chosen sites of the larger host lattice shall
represent the positions of the Ru particles.

Following previous simulations \cite{Vi89,Me83}, we then add
spherical Au particles one at the time undergoing a
random walk starting from a point on a variable circle centered on the
intersection of both squares (containing the central seed particles).
As a difference, however, in our case
the radii of such circles change according to the distance
from the squares' center to the most distant particle in the (outer)
growing islands in which several (Ru) seed particles are also contained.
If in this process the Au particles, that are being
deposited adjacent to (occupied Au or Ru) lattice sites, reach
a point that is larger than a few times the diagonal of the smaller square,
then these are not longer included.  We started off again the random walk
at a nearest randomly chosen position remembering that the
variable circles enclose the growing clusters untill eventually they
cover all thirteen.
This means that islands are only allowed to growth gradually
starting from the centre of the square lattice for simulations
towards the corners of the square containing the outer four
(Ru) seed particles.

Although this stochastic model we are proposing for modelling
the present phenomena may be seen as heuristic,
we shall show next that, in fact, by using this algorithm
the fractal growth of all inner islands contained around
the central one becomes essentially `simultaneous'.
In our new application of the DLA model to the fractal growth
of Au on Ru(001), we shall also see below that sticking
probabilities less than unity (correponding to DLA) on island
formation will allows to reproduce nicely the annealing effects
as seen experimentally.
The radius of gyration for each island will be here related
to the number of particles contained in such single clusters by \cite{Me83}
\begin{equation}\label{eq:1a}
R^{(n)}_{g}\sim N^{1/d_{f}}
       \;\;\;\;\;\;  n=1,2,3, \cdots  \;\;\; ,
\end{equation}
where $d_{f}$ is the Hausdorff (fractal) dimension of the $n$-island
having a size which is large enough.

We now focus on the results of our stochastic {\em ad-hoc} DLA version.
An example of the structures generated by computer is shown in Fig.1.
This figure shows 13 irregular 2D dentritic islands of (Au) particles that
gradually grow on a larger host lattice of (Ru) particles.  The total number
of deposited (Au) particles is 2200 in a $260\times 260$ square lattice.
The islands contain more than 100 particles each.
The (different) arrival times for the first 1100 added particles are
represented in Fig.1 using open dots.
This distintion also enables to show that the fractal growth of all inner
islands contained around the central one becomes almost `simultaneous'.

Model particles are added one at the time to a particular DLA island,
that are contained well inside the edges of the simulation box,
via random walk trajectories that have originated from the fluctuating
inner region compressed between the central growing cluster and
the corresponding square edges (also shown in Fig.1).
Let us see below how these irregular island shapes may be removed from
their thermal stability.

In order to reproduce annealing effects on the clusters morphology,
Fig.2 shows the results of our
DLA-based model simulation of the 13 irregular dentritic islands
as in Fig.1 but assuming a sticking probability of $s\approx 0.17$ at
nearest-neighbour positions only.
These new islands contain more than 400 particles each and in total
sum up 8600 particles.
The arrival times in this case for the first half of added particles is
also represented in Fig.2 by full dots.

It is evident from Fig.2 that different sticking
probabilities on island formation lead to reproduce smoother structures
as seen experimentally \cite{Hw91}.  Since we have assumed for
simplicity that there is no mass transfer between Au islands more extended
and connected structures can not be formed.  This suggest that
these denditric forms growing with a sticking probability different from 1
({\em i.e.} DLA) are related to the diffusion of Au particles not being
stick on each contact.  As in the case of single clusters
reported in the literature \cite{Me83}, in our study a small sticking
probability also leads to generate denser islands.

We examine next the fractal properties of the simulated fractal islands
displayed in Figs.1 and 2.  We count the number of
particles $N(r)$, {\em c.f.} Eq.(\ref{eq:1a}), inside a circle of
increasing radius $r$
(in lattice units) around each (Ru) seed particle and plot it as a
function of $r$ in a log-log plot as depicted in Fig. 3(a), (b) and (c).
These curves corresponds to Au particles inside the
circles centered at (a) the four corners
of the outer squares in Figs.1 and 2 (drawn with solid lines),
(b) the center of the four edges
of the outer squares in Figs.1 and 2 (drawn with solid lines),
(c) the four corners
of the inner squares in Figs.1 and 2 (drawn with dashed lines).

For comparison, in Fig.3 the functions $N(r)$ for the central islands along
the intersection of the diagonals of both squares in Figs.1 and 2 are
also plotted (using connected dots) throughout 3(a), 3(b) and 3(c).
These results have been obtained by adding up to about 1500
particles to the single
central clusters in the absence of any other seed (Ru) particle.
It can be immediately concluded that, to a good approximation, the simulated
2D dentritic islands reveal DLA fractal dimension $d_{f}\approx 1.70$
\cite{Wi81,Ca91,Ca91a} for almost one decade in Fig.3.
The central DLA cluster of Fig.1 leads to
lower values of $\ln N(r)$ than the central DLA cluster of
Fig.2 for a fix value of $\ln r$ but both slopes are found to be parallel.
These interesting findings encourages us to give a further justification
to our simulation model which allows the fractal islands to growth gradually
from the centre of the simulation box towards the four corners of the square
containing the outer (Ru) seed particles.

Using the present algorithm for the stochastic simulation,
it becomes thus possible to generate highly irregular 2D
dentritic islands in reasonable accord with observations of Ref.\cite{Hw91}.
Fractal dimensionality in such set of complex patterns
{\em all} present DLA-$d_{f}$.  This is important considering that
13 irregular islands grow almost `simultaneously'.  These results
seem to reflect the essential physics behind these complex phenomena,
namely, the growth being dominated by Au particles migrating via random
walks over the Ru surface. Since contributions to the random growth process
from the spherical (Au) particles in the central region
become extremely small on succesive increasing of the amount of
fractal islands, this may -to some extent- be a way of taking
into account the presence of kinetic limitations existing at room
temperature in thin-film growth.  In fact, further insight
on changes of the average island radius with `time' can be obtained
as discussed next.

For 2D random walks, the mean squared distance made by particles during a
given time interval $t$ satisfies $<r^{2}(t)>\sim t$.  Therefore, when
measuring the extend of Brownian trajectory by the total number of places
visited by a particle making $t$ steps, the above is equivalent to \cite{Vi89}
\begin{equation}\label{eq:2a}
N(r) \sim <r^{2}(t)> \sim t  \;\;\;  .
\end{equation}
This relationship between $t$ and the number of particles
covering a circle of radius $r$, enables us to analyse the islands
growth in analogy to diffusion-limited, precipitate growth with soft
impingement of precipities \cite{Lev91}.  In drawing this analogy, we follow
Ref.\cite{Lev91} and consider immobile Au-islands growing on Ru(001) as the
precipitates.  As a difference, in this work we concern ourselves
with experiments on fractal growth of 2D Au islands on Ru(001) \cite{Hw91},
whereas in Ref.\cite{Lev91} Au-atoms were deposited on glass substrates forming
(rather circular) island films.  In both measurements, however, evidence of
island inmigration processes (or mobility) has been found upon annealing.

Assuming the area in the island migration process to be conserved,
then thin film island growth kinetics can -to a crude approximation- be
described in terms of post-deposition island growth.
Thus, the growth of nucleated precipitates can be described by
the stretched exponential respose: \cite{Lev91}
\begin{equation}\label{eq:3a}
(\frac{r(t)}{r_{o}})^{3} =1+R[1-e^{(-t/T_{o})^{m}}]   \;\;\; ,
\end{equation}
where $r(t)$ is an average island radius at time $t$ and $r_{o}$ is
the initial radius, $T_{o}~\propto~e^{+Q/k_{B}T}~/~D_{o}$
is a time constant proportional to the island diffusion coefficient
$D_{o}$ and migration activation energy $Q$, and $R$ is the
fractional island size increase at $t\rightarrow \infty$.
In the above the dimensionless exponent $m$ equals 1 because the
random walk diffusion of single particles (until they reach the growing
fractal clusters) is confined to 2D.  So, in the present case, the
stretched exponential relaxation of Eq.(\ref{eq:3a}) does not differ
from a simple exponential law.   For post-deposition island growth,
surface diffusion is confined to 2D and island growth occurs in
3D \cite{Lev91}.

In order to give a simple estimate of the `time' variations of
the average island radius we use next the results of our multiple DLA growth
model, ({\em c.f.} Eq.(\ref{eq:1a})), in conjuction
with relations (\ref{eq:2a}) and (\ref{eq:3a}).   In doing this,
we have that islands execute Brownian motion at a given
temperature and time.  Figure 4 shows results obtained using our
simulation data for the central irregular 2D-DLA island of
Fig.1.  Similar results are found when considering the
12 nearest islands or, alternatively, those of Fig.2.  This is because,
as we have shown in Fig.3, the slopes are to a good approximation
{\em all} parallel.

As a result of island mobility due to atomic motions in the islands
it can be shown that the relation $r(t)/r_{o}=h_{pl}/h_{p}(t)$ holds,
where $h_{pl}$ is the initial peak position of scattering
patterns and $h_{p}(t)$ is the position at a given succesive
time \cite{Lev91}.  Then, we are able to predict that
in the fractal growth of Au films deposited on Ru(001),
there should be a main peak in (small angle) scattering intensity data due
to the fact that the average center-to-center island spacings are of the
order of average island sizes.  This phenomenon being also observed in
the circular islands growth of Au-atoms deposited on glass substrates.
{}From the curves in Fig.4, that characterize the growth and post-deposition
`kinetics' of our simulated Au-thin film islands on Ru, we also
argue that there should be a rapid increase in island sizes followed by a
plateu region.  As a first order approximation, this plateu should decrease
when increasing the migration activation energy assuming the single atom
diffusion coefficient $D_{o}$ to be constant in 2D.

In this way we have taken the first steps towards understanding
multiple fractal growth of 2D islands.
A simple model based on irreversible DLA has been introduced
inspired by the properties of fractal growth of Au films deposited on Ru(001).
Discrete 2D dentritic islands of Au atoms that
grow on a larger host lattice of Ru atoms, having DLA fractal dimension each,
were generated via an extended version of the DLA algorithm for single
aggregates.  Annealing effects on the islands morphology
were reproduced assuming different sticking probabilities at
nearest-neighbour lattice sites.  From these simulations, the islands
growth were described in analogy to diffusion-limited,
precipitate growth with soft impingement of precipities.  This allowed us to
analyse thin island growth `kinetics' of Au films on Ru(001)
and predict a main peak in (small angle) scattering intensity pattern.
As a main difference with respect to gold
islands on glass substrates \cite{Lev91}, fractal growth of Au films on Ru(001)
is dominated by particles migrating (only) via random walks over a surface of
several seed obstacles.  In addition, we may also make variations of the model,
{\em e.g.}, to include random (Ru) defects or generate a larger
amount of islands, but to achieve this larger scale simulations are needed.

\section*{Acknowledgements}

The authors acknowledge the Condensed Matter Group
and the Computer Centre at ICTP, Trieste, for valuable support.
Sincere thanks are also extended to the referee for pointing out
important comments.

\newpage

\section*{Figure captions}

\begin{itemize}

\item {\bf Figure 1}:
DLA-based model simulation of 13 irregular 2D dentritic islands
of (Au) particles that gradually grow on a larger
host lattice of (Ru) particles.

\item {\bf Figure 2}:
DLA-based model simulation of 13 irregular dentritic islands
as in Fig.1 but assuming a sticking probability of $s\approx 0.17$ at
nearest-neighbour positions only.

\item {\bf Figure 3}:
Number of spherical particles $N(r)$ inside a circle of radius $r$
centered at (a) the corners of the squares in Figs.1 and 2 (full lines),
(b) the edges of the squares in Figs.1 and 2 (full lines),
(c) the corners of the squares in Figs.1 and 2 (dotted lines).
The $N(r)$ values for the central islands, lying on
the intersection of the square diagonals of Figs.1 and 2, are plotted
by (connected) black dots throughout (a), (b) and (c).

\item {\bf Figure 4}:
Exponential respose of the reduced average island radius
$[(r(t)~/~r_{o}~)^{3}~-~1~]~/~R$ at `time' $t$ and
different time constant $T_{o}$ proportional to
the island diffusion coefficient and migration activation energy.

\end{itemize}

\end{document}